\documentclass[twocolumn, times]{aastex631}

\usepackage{amsmath, latexsym, color, verbatim}

\shorttitle{The Highly Ionized Wind of 10 Lac}
\shortauthors{Law et al. 2024}

\newcommand{\kms}{\textrm{km~s}\ensuremath{^{-1}\,}}

\newcommand{\nevi}{\textrm{[Ne\,{\sc vi}]}}
\newcommand{\nev}{\textrm{[Ne\,{\sc v}]}}

\newcommand{\ofour}{\textrm{[O\,{\sc iv}]}}

\def\one{\,{\sc i}}             % for producing Na I as Na\one\ etc.
\def\two{\,{\sc ii}}
\def\three{\,{\sc iii}}
\def\four{\,{\sc iv}}

\begin{document}

\title{JWST/MIRI detection of \nev, \nevi, and \ofour\ wind emission in the O9V star 10~Lacertae}

\author[0000-0002-9402-186X]{David R.\ Law}
\affiliation{Space Telescope Science Institute, 3700 San Martin Drive, Baltimore, MD, 21218, USA}

\author[0000-0003-0145-8964]{Calum Hawcroft}
\affiliation{Space Telescope Science Institute, 3700 San Martin Drive, Baltimore, MD, 21218, USA}

\author[0000-0002-0806-168X]{Linda J. Smith}
\affiliation{Space Telescope Science Institute, 3700 San Martin Drive, Baltimore, MD, 21218, USA}

\author[0000-0003-2429-7964]{Alexander W. Fullerton}
\affiliation{Space Telescope Science Institute, 3700 San Martin Drive, Baltimore, MD, 21218, USA}

\author[0009-0005-7288-6407]{Christopher J. Evans}
\affiliation{European Space Agency (ESA), ESA Office, Space Telescope Science Institute, 3700 San Martin Drive, Baltimore, MD 21218, USA}

\author[0000-0001-5340-6774]{Karl D.\ Gordon}
\affiliation{Space Telescope Science Institute, 3700 San Martin Drive, Baltimore, MD, 21218, USA}

\author[0000-0002-5320-2568]{Nimisha Kumari}
\affiliation{AURA for European Space Agency (ESA), ESA Office, Space Telescope Science Institute, 3700 San Martin Drive, Baltimore, MD,
21218, USA}

\author[0000-0003-2685-4488]{Claus Leitherer}
\affiliation{Space Telescope Science Institute, 3700 San Martin Drive, Baltimore, MD, 21218, USA}

\begin{abstract}

We report the detection of broad, flat-topped emission in the fine-structure lines of \nev, \nevi, and \ofour\ in mid-infrared spectra of the O9 V star 10 Lacertae obtained with JWST/MIRI.  Optically thin emission in these high ions traces a hot, low-density component of the wind. The observed line fluxes imply a mass-loss rate of $>3$\,$\times\,$10$^{-8}$\,M$_{\odot}$\,yr$^{-1}$, which is an order of magnitude larger than previous estimates based on UV and optical diagnostics. The presence of this hot component reconciles measured values of the mass-loss rate with theoretical predictions, and appears to solve the ``weak wind'' problem for the particular case of 10 Lac.

\end{abstract}


%%%%%%%%%%%%%%%%%%%%%%%%%%%%%%%%%%%%%%%%%%%

\section{Introduction}

Mass loss via stellar winds is a critical component of the evolution of massive stars, but our understanding of such winds is incomplete.  While theoretical models successfully predict the mass outflow rates from the most massive O-type stars, a sudden reduction in wind strength is observed below a threshold luminosity of 
log$(L/L_{\odot} ) \sim 5.2$.
This so-called `weak-wind problem' has been 
confirmed in multiple star-forming environments in the SMC \citep{bouret03, martins04}, LMC \citep{brands22, hawcroft24}, and the Milky Way \citep{martins05, marcolino09, dealmeida19} with a drop in observed mass-loss rates of around 2 orders of magnitude compared to theoretical predictions.
Indeed, below this threshold luminosity all typical wind diagnostics become very weak or disappear completely, meaning that only upper limits on the mass-loss rates can be set. 

It remains unclear whether the winds of massive stars in this low luminosity regime are truly so weak (which would impact evolutionary and feedback predictions), or whether 
they may simply not be visible in the traditional UV/optical wavelength regimes.
If the winds are genuinely weak, theoretical solutions have been proposed suggesting
a possible loss of line-driving force from the recombination of iron \citep{vink22}, a decoupling of minor ions and the bulk winds \citep{owocki02}, or simply an incomplete line-driving formalism \citep{lattimer21}. Alternatively, since the low density outflows in O dwarfs result in inefficient cooling, it is possible that the winds from late-type O stars could be shock heated to extremely high temperatures and diagnostics would thus be available at only higher ionization stages \citep{lucy12, lagae21}.  Such higher ionization stages would only be observable in the extreme-UV or X-ray regime, or in the mid-infrared via fine-structure lines.

10~Lacertae (10 Lac, HD\,214680) is one of the most well-studied late O-type dwarfs, with a low projected rotational velocity of $v$ sin$(i)$\,$=$\,14\,\kms and a macroturbulent broadening of 43\,\kms\ \citep{holgado22}.
Given its narrow lines and bright visual magnitude (V = 4.9), 10~Lac is one of the primary classification standards for the O9~V type \citep{sota11} and has often been used as a reference target in quantitative studies of O-type stars.
Likewise, 10 Lac is a
well established weak-wind star. Early mass-loss rate estimates from H$\alpha$ were not able to identify the low wind strength and set an upper limit of $\sim\,$10$^{-7}$\,M$_{\odot}$\,yr$^{-1}$ \citep{leitherer88}. An additional, independent upper limit on the mass-loss rate of $\sim\,$10$^{-7}$\,M$_{\odot}$\,yr$^{-1}$ comes from a non-detection in the radio \citep{bieging89, vallee85,schnerr07}. UV analysis was only able to set a maximum wind velocity of $1070\pm50$\,\kms 
from \ion{N}{5} \citep{prinja90}. This wind velocity was further supported by the UV fit from \citet{martins04} %[\textcolor{blue}{is this the correct reference - this paper is on SMC N81 stars}] [\textcolor{purple}{Yes they fit 10 lac to compare to their SMC sample}]
but the mass-loss rate upper limit was revised down to $\sim\,$10$^{-8}$\,M$_{\odot}$\,yr$^{-1}$ in \citet{herrero02} with these authors and \citet{martins04} using $\sim\,$10$^{-10}$ and $\sim\,$10$^{-9}$\,M$_{\odot}$\,yr$^{-1}$ in their respective best-fits.

Here we report the discovery of highly ionized, broad forbidden emission lines arising in the outer wind of 10 Lac, suggestive of an unusually hot wind component. Such broad, flat-topped forbidden fine-structure lines have been observed before in the mid- and far-IR spectra of 
Wolf-Rayet stars such as $\gamma^2$ Velorum  in the ionic species of [S\four], [Ne\two], [Ne\three] and [O\three] \citep{barlow88, crowther24}. As shown by \citet{barlow88}, the rectangular line profiles indicate that these features are optically thin and formed in the outer parts of the stellar wind, where the densities are low and the wind has reached its terminal velocity. Because of their simple formation conditions, the lines can be used as diagnostics to determine stellar properties such as abundances and mass loss rates.

%first concrete evidence for a significant mass outflow from 10 Lac, as traced in extremely highly ionized, low density gas in the mid-infrared.  

We describe the observations and data processing techniques in \S \ref{obs.sec}, present the MIRI spectra and measurements in \S \ref{results.sec}, and discuss the implications of these results in \S \ref{discussion.sec}.

\section{Observations and Data Processing}
\label{obs.sec}

We observed 10 Lac in July 2022 using the Mid-Infrared Instrument's (MIRI) Medium Resolution Integral Field Unit Spectrometer \citep[MRS;][]{argyriou23} onboard the James Webb Space Telescope (JWST) as part of a Cycle 1 calibration program
designed to refine the MRS geometric distortion solution (JWST Program ID [PID] 1524, PI: Law).
This program selected 10 Lac as the target for its brightness, rich parallel imaging field for measuring absolute astrometry, and potential utility in characterizing
the MRS fringe and photometric calibration due to its smooth and well characterized spectrum and
lack of variability \citep[see, e.g.,][]{gordon22}.

A custom 57-point dither pattern was used to dither the source widely around the field of view.  With an exposure time of 55 seconds per exposure to maximize the S/N ratio at long wavelengths, the total exposure time was thus approximately 0.9 hours per spectral band. 
In addition to serving as the basis of the current MRS distortion solution \citep{patapis24}, these observations of 10 Lac also underpin
the MRS photometric calibration shortward of 18 $\micron$ \citep{law24}.

Even deeper observations of 10 Lac were obtained as a part of calibration
PID 3779 (PI: Gasman) designed to assess the impact of sub-pixel
positioning on the spectral fringing of the MRS \citep{gasman23}.  This program obtained a 9-point raster scan
around each of 8 primary dither locations, for a total of 72 exposures
and an effective integration time of 5.5 hours per spectral band.  These data were obtained in November 2023,
approximately 16 months after the data from PID 1524.

Both sets of data were processed through the standard {\sc calwebb\_detector1}, {\sc calwebb\_spec2}, and {\sc calwebb\_spec3} stages of the JWST calibration pipeline \citep{bushouse24} using the
default STScI pipeline notebook for the MIRI MRS observing mode\footnote{https://github.com/spacetelescope/jwst-pipeline-notebooks}.  We used the software and calibration reference files
corresponding to pipeline Build 11.0; i.e., {\it jwst} software version 1.15.1 and calibration reference
data context 1276.
We used the default spectral extraction aperture with radius $r = 2$ times the PSF FWHM in Channels 1--3 (i.e., shortward of 18 $\micron$), but set $r = 1$ FWHM in Channel 4 (longward of 18$\micron$) in order to minimize the impact of annular background subtraction errors on the source spectrum which is faint at these long wavelengths.  We additionally included the optional pipeline steps that applied bad pixel replacement and a 1-d residual fringe correction to the extracted spectra \citep{kavanagh24}.

While 10 Lac is classified as a double/multiple star, the first companion is 5 magnitudes fainter in the optical and separated by approximately 1 arcminute, while the second companion is about 10 magnitudes fainter than 10 Lac and separated by 3.6 arcseconds \citep{mason01}.
Both companions lie outside the FOV of the MRS.

\section{Results}
\label{results.sec}

\begin{figure*}[!htbp]
\epsscale{1.15}
\plotone{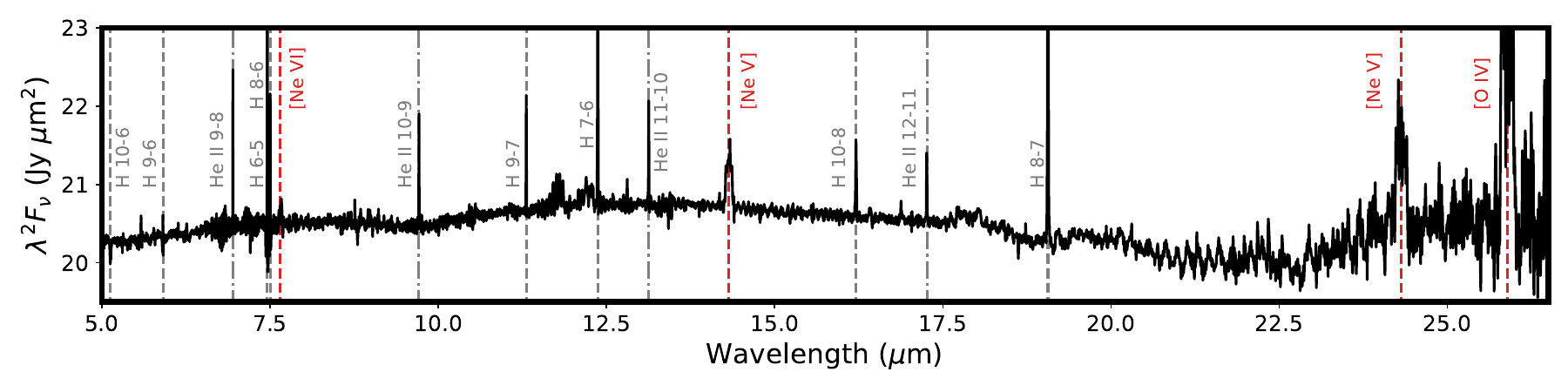}
\caption{MIRI MRS spectrum of 10 Lac from PID 1524 plotted in Rayleigh-Jeans units.  Dashed and dot-dashed vertical gray lines respectively indicate H\one\ and He\two\ photospheric emission features due to non-LTE effects, dashed red lines indicate highly ionized broad stellar wind \nevi, \nev, and \ofour\ transitions.  The extremely broad depression around 10$\micron$
is a known silicate dust absorption feature.
%\textcolor{blue}{Include or omit [Ne II]?}
}
\label{fig1.fig}
\end{figure*}
\begin{figure*}[!htbp]
\epsscale{1.17}
\plotone{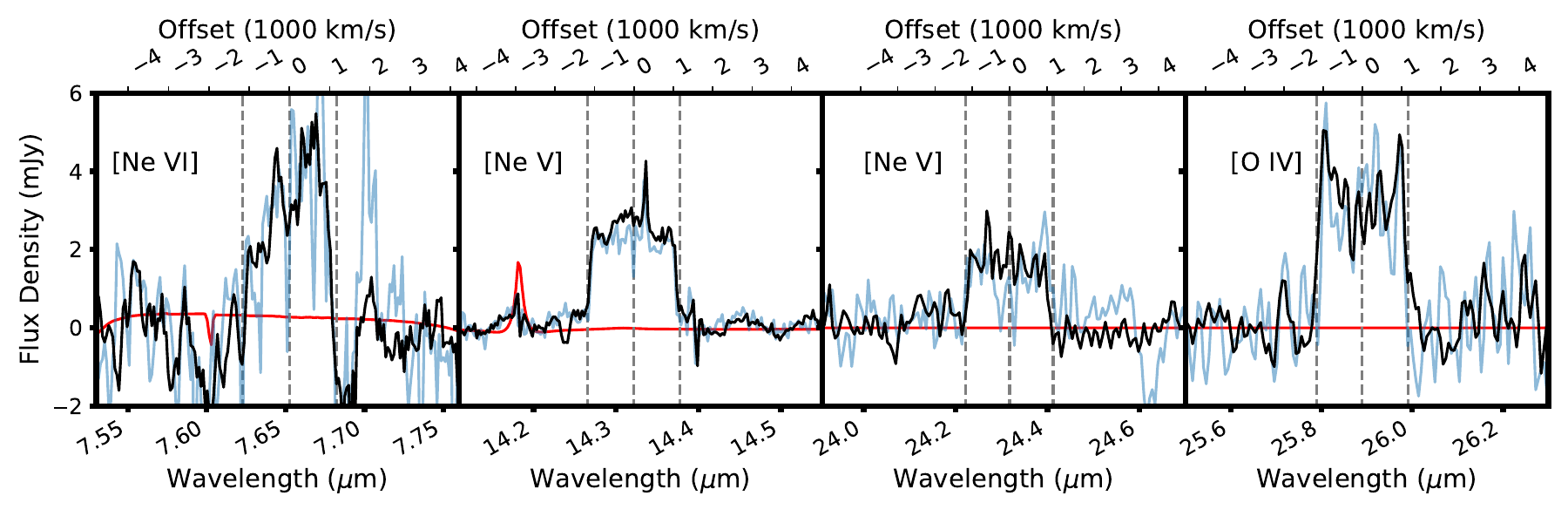}
\caption{Continuum-subtracted spectrum of 10 Lac from PID 1524 (solid black line), zoomed in around the wavelengths
of the broad \nevi, \nev, and \ofour\ features.  The light blue line shows the spectrum observed 16 months later by PID 3779 for comparison.  The solid red line represents the corresponding continuum-subtracted CALSPEC model; the emission feature at 14.183 $\micron$ is H\one\ 13-9.  Dashed vertical lines indicate $\pm 1170$ \kms\ around the rest wavelength of each line.
}
\label{fig2.fig}
\end{figure*}

Figure \ref{fig1.fig} shows the 
5--26.5 $\micron$ spectrum of 10 Lac from program 1524 in Rayleigh-Jeans units.
While some high-frequency artifacts from residual fringing are evident (e.g., around 12 $\micron$), the spectrum is generally smooth with a typical S/N per wavelength bin ranging from $\sim$ 1000 at $\lambda = $5--18 \micron\ to $\sim 200$ at $\lambda = 22 $\micron\ and $\sim 30$ longward of $\lambda = 25$\micron.

At the spectral resolving power of the MRS \citep[$R \sim 4000$ at 5 $\micron$, $R \sim 1500$ at 25 $\micron$;][]{argyriou23} we are able to detect numerous narrow H\one\ and He\two\ emission features.
These features have a mean velocity of $-7.0\pm7.9$\,\kms\ (consistent with the 10 Lac systemic velocity of -10 \kms) and a kinematic width of $\sigma_{\rm HI} = 42 \pm 5$ \kms\ after correction for the instrumental line spread function.
These lines are %unresolved and are 
formed due to non-LTE effects in the stellar photosphere \citep{najarro11}.  
Both H\one\ and He\two\ emission lines are relatively weak, with most peaking at 10\% or less of the continuum flux (with the exceptions
of the H\one\ 6-5, 7-6, and 8-7 transitions) and would have been difficult to detect at the $R \sim 600$ spectral resolution characteristic of prior Spitzer IRS observations \citep[e.g.,][see their Fig. A3]{marcolino17}.

The most conspicuous features in the 10 Lac spectrum are four broad emission lines that we identify
as \nevi, \nev, and \ofour\ (see Table \ref{results.table}).  These highly-ionized species have ionization potentials of up to 126 eV for Ne$^{5+}$ suggestive of high gas temperatures.
We show continuum-subtracted spectra zoomed in around each of these features in Figure \ref{fig2.fig},
along with the continuum-subtracted CALSPEC \citep{bgt14, bohlin22} model spectra, which contain only photospheric features.
The spectra obtained from both PID 1524 and 3779 are consistent to within
observational uncertainty.
Despite the shorter integration times, the effective S/N ratio is higher in the PID 1524 spectra, likely because the 57-point dither marginalizes over flatfield systematics that may limit performance in PID 3779.
Combining the data from both programs, we report integrated fluxes for each of these four lines in Table \ref{results.table}.

% Standard stars
\begin{deluxetable*}{cccccc}
\label{results.table}
\tablecolumns{6}
\tablewidth{0pc}
\tabletypesize{\small}
\tablecaption{Parameters of Forbidden Emission Lines Detected in 10 Lac}
\tablehead{
\colhead{} & \colhead{Rest Wavelength} & \colhead{Ionization} & \colhead{Flux} & \colhead{Critical Density\tablenotemark{a}} & \colhead{Formation Radius}
\\
\colhead{Line} & \colhead{(\micron)} & \colhead{Potential (eV)} & \colhead{(10$^{-15}$ erg s$^{-1}$ cm$^{-2}$)}
& \colhead{(log($n_e$/cm$^{-3}$)}) & \colhead{(R$_*$)}
}
\startdata
\nevi & 7.6524 & 126 & $10.1 \pm 0.5$ & 6.0 & 60\\
\nev & 14.3217 & 97 & $3.80 \pm 0.05$ & 5.2 & 230\\
\nev & 24.3175 & 97 & $1.37 \pm 0.06$ & 5.0 & 275\\
\ofour & 25.890 & 55 & $2.94 \pm 0.06$ & 4.2 & 490\\
\enddata
\tablenotetext{a}{Assuming $T_e$ = 35 kK.}
\end{deluxetable*}

The \nev\ lines in particular exhibit flat-topped profiles commonly seen in 
lower-ionization species\footnote{E.g., [S\four] 10.5 \micron, [Ne\two] 12.8 \micron, [Ne\three] 15.5 \micron\ with ionization potentials of 35 eV, 22 eV, and 41 eV respectively.} (which are not detected here) in Wolf-Rayet stars
\citep[][]{barlow88,crowther24}.
Following \citet{barlow88}, we measure the half-width at zero intensity and find an outflow speed
of $1170 \pm 50$ \kms.  This value agrees with the 
terminal wind velocity of 1070 \kms\ estimated by previous studies of the UV spectrum \citep{prinja90, martins04}.
The \nevi\ and \ofour\ lines in contrast show evidence of structure, with \ofour\ potentially being double-peaked
with ``horns'' at $\pm \, 970$ \kms\ with respect to the systemic velocity.
As discussed by \citet{ignace06},
the flat-topped profiles are indicative of a constant-velocity spherical stellar wind, while substructure
in these profiles may be related to asymmetric winds and differences in the geometry of gas with different ionization energies (with horned profiles in particular resulting 
when the stellar wind is denser either at the poles or the equator).

%We also detect tentative evidence for narrow and weak \neii\ $\lambda$ 12.8135 \micron\ emission.  This feature is significantly contaminated by residual fringing artifacts and is unresolved. \textcolor{blue}{Should we mention this or remove as it is tentative?}

\section{Discussion}
\label{discussion.sec}

\begin{figure}%[!htbp]
%\begin{figure}[t!]
%\epsscale{1.15}
%\plotone{10lac_ne.pdf}
\includegraphics[width=\columnwidth]{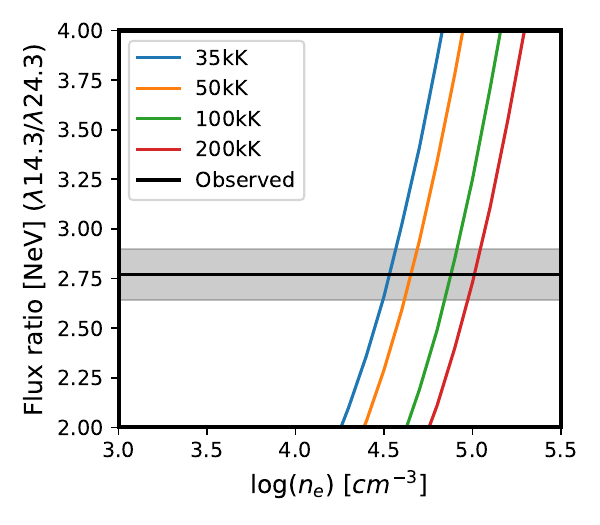}
\caption{Flux ratios for \nev\ 14.3 \micron\ / 24.3 \micron\ as a function of the electron temperature $T_e$ and density $n_e$.  Colored lines show model estimates produced using EQUIB \citep{howarth81, danehkar20} for values of $T_e$ in the range 35-200\,kK.  The solid black line and grey shaded region show the measured flux ratio and 1$\sigma$ uncertainty for 10 Lac, respectively.}
\label{fig3.fig}
\end{figure}

In order to better understand the physical conditions in the stellar wind we aim to constrain the electron temperature ($T_{e}$) and density ($n_{e}$) in the fine-structure line-forming region.
%It is difficult to constrain $T_{e}$ with the ratio of the observed IR fine structure line strengths\footnote{The larger energy level separation of \nev\ 3426\ \AA/24.3 \micron\ is more sensitive to temperature, but we would not expect to detect such faint emission atop the 70~Jy stellar continuum of 10 Lac at near-UV wavelengths.} as these transitions are dominated by electron-ion collisions making their relative strengths dependent on both $T_{e}$ and $n_{e}$, as opposed to radiative decay transitions which depend only on $T_{e}$.
  The \nev\ 14.3 \micron\ / 24.3 \micron\ ratio is most sensitive to density across a wide range of temperatures \citep[e.g.,][]{sturm02}.
Assuming a broad range of $T_e = $35-200\,kK (with a lower limit of 35 kK set at the effective temperature of 10 Lac from \citealp{aschenbrenner23}), the observed \nev\ 14.3 \micron\ / 24.3 \micron\ ratio of $2.77 \pm 0.13$ corresponds to a density range log($n_e$/cm$^{-3}$) $= 4.5$ -- $5.0$, as shown in Figure \ref{fig3.fig}.  This range is broadly consistent with upper limits placed on $n_e$ by the critical density\footnote{At the critical density, spontaneous emissions equal collisional de-excitations for a given energy level.} $n_{c}$ for the transitions. These are listed in Table \ref{results.table} for $T_{e}=35\,$kK; all of these values become much larger at higher electron temperatures.

%  , while the larger energy level separation of \nev\ 3426\ \AA/24.3 \micron\ is more sensitive to temperature. HST/STIS spectra of 10 Lac, however, show no evidence of \nev\ 3426 \AA\ emission but we would not expect to be able to detect such faint emission atop the 70~Jy stellar continuum of 10 Lac at near-UV wavelengths.  The non-detection of \nev\ 3426 \AA\ thus provides no meaningful bounds
%on the $T_{e}$ for the \nev\ 3426 \AA/24.3 \micron\ line ratios of $\sim$ unity or below
%predicted by the PyNeb modeling tool \citep{lurid15}. 
%\textcolor{red}{If we're discussing the optical spectra should we not mention the MIMES or IDS data? Maybe just remove discussion of optical? @Chris / Alex}  [\textcolor{blue}{Should we just omit the discussion on the optical [Ne\five] line - it does not really add anything to the paper]}

Taking the equation of mass continuity as in \cite{smith05} with $n_{c}$ and assuming $\dot{M}=3$\,$\times\,$10$^{-8}$\,M$_{\odot}$\,yr$^{-1}$ (see below), we can calculate the radius at which the lines are formed assuming a smooth radial density profile; these are listed in Table \ref{results.table} in terms of the stellar radius (R$_*$ = 7.4 R$_{\odot}$, \citealp{aschenbrenner23}).
%These radii assume a
%{\bf smooth radial density profile, and increase for clumpy profiles.}
% Calum (since you're editing too!) is this accurate? I would just remove all mentions of clumping before the discussion, but it is accurate.  Sounds good.
%clump filling factor $f_{cl}$ of 1 and increase as $f_{cl}$ is reduced. 
We find that all the lines are formed below the critical density, at radii that are consistent with the wind having reached terminal velocity given a standard $\beta$ acceleration profile.

%Can we say something based on the comparative flux between the lines?  Need to measure this.
%\textcolor{magenta}{Chris: Artemio Herrero might have an old WHT spectrum that covers the blue ~3400\AA\ region; I've asked if he is happy to share it so we can inspect the blue Ne lines. There are also some blue EMMI data in the ESO archive from 2002 that Gregor Rauw was a co-I on.} 
%\textcolor{purple}{It is difficult to constrain the electron temperature with the ratio of observed \nev\ lines as these transitions are dominated by electron-ion collisions making their relative strengths dependent on both $T_{e}$ and $n_{e}$, as opposed to radiative decay transitions which depend only on $T_{e}$. We can perhaps gain more insight by quantifying the non-detection of other lines. We confirm a non-detection of \nev\ lines in the optical with S/N = .. using archival spectra .... We also find the IR \nev\ lines are predicted to be much stronger than the optical \nev\ line ($\lambda$ = 3447\AA), by a factor of 300 at the critical density of \nev\ 14.3 $\micron$ and $T_{e}$=10kK, with the ratio increasing with lower $T_{e}$ or at lower densities (computed using \textsc{Equib} from \citealp{howarth81}). We can place broad upper limits on $T_{e}$ given the ratio decreases above 10kK, at 100kK the ratio flips and \nev\ 3447\AA\, is predicted to be stronger than the IR lines. }

% Ions are very highly ionized, but electrons will generally have a much
% lower temperature.  What should we assume for it?

IR forbidden fine-structure emission lines have often been used to determine abundances in %planetary nebulae and 
the stellar winds of Wolf-Rayet stars, given that line strengths depend mainly on the wind strength and abundance \citep{barlow88, dessart00, ignace01, morris04, smith05, crowther24}.  
The ionic abundances $\gamma_{i}$ of these types of line transitions depend on the mass-loss rate ($\dot{M}$) as $\gamma_{i} \propto \dot{M}^{-1.5}$. Therefore the lines we observe in the outer wind of 10~Lac can be used to put limits on $\dot{M}$ for a given line flux $I_{ul}$,  assumptions for $T_{e}$, $n_{e}$ and  knowledge of the distance and wind speed. We follow the methodology of \citet{barlow88}, \citet{dessart00} and \citet{crowther24} and provide additional details, including the full expressions, in the Appendix.

\begin{figure}%[t!]
    \includegraphics[width=\columnwidth]{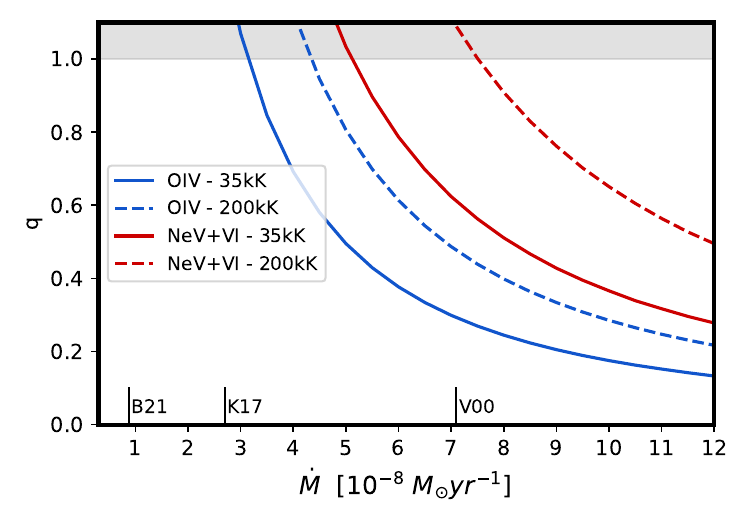}
    \caption{Ionisation fraction $q$ as a function of mass-loss rate for O$^{3+}$ (blue) and Ne$^{4+}$ + Ne$^{5+}$ (red) ions. Solid and dashed lines show electron temperatures of $T_{e} = $ 35 and 200 kK, respectively. Values of $q = 1$ represent 
    the lower limit of the mass-loss rate for each curve assuming that the indicated ionization states comprise the total contribution of the respective element in the line-forming region.  Values of $q > 1$ (grey shaded region) are thus unphysical, and any unaccounted-for species will drive $q < 1$. Vertical black ticks show theoretical mass-loss rate predictions from \cite{bjorklund21}, B21, \cite{krticka17}, K17 and \cite{vink00}, V00.}
    \label{fig.mdot}
\end{figure}

%\begin{figure}%[t!]
    %\includegraphics[width=\columnwidth]{mdot_abundance_updated.png}
%    \includegraphics[width=\columnwidth]{mdot_abundance_nev_10lac_paper_v2.pdf}
%    \caption{Neon (red) and Oxygen (blue) abundances as a function of mass-loss rate. Solid and dashed lines show $T_{e}$ of 10 and 100kK for Ne and 10 and 50kK for O. Vertical error bars are representative of the uncertainty about each of the abundance curves if the clumping factor is varied. Horizontal red and blue ticks show measured abundances from the literature; \textbf{where these cross the abundance curves we plot circles to guide the eye to the implied mass-loss rates.}  Vertical black ticks show theoretical mass-loss rate predictions from \cite{bjorklund21}, B21 and \cite{krticka17}, K17.}
%    \label{fig.mdot}
%\end{figure}

% upon the ionic abundance of the transition which can only be constrained if the wind strength is known, this is fine for WR stars, or even typical O stars where the wind strength (clumping corrected) can be determined from optical and UV analysis. Determining abundances for 10 Lac then becomes difficult given the uncertainty of the wind strength. On the other hand, if the abundance is known, the forbidden line should give an estimate of the wind strength. \textcolor{blue}{LJS: What about the uncertainty for the electron temperature?}
In Figure \ref{fig.mdot} we show predictions for the ionisation fraction $q$ (i.e., the fraction of atoms of a given element in a particular ionization state) using the measured line fluxes over a range of input mass-loss rates. We find that the trends in $q$ with $\dot{M}$ depend on $T_{e}$, but we can place lower limits on the mass-loss rate on the assumption that the entire neon and oxygen mass budget is in the observed 
Ne$^{4+}$, Ne$^{5+}$, and O$^{3+}$ ions.
%entire Neon and Oxygen mass is contas the lowest possible allowed mass-loss rate is reached when the total %contribution for any element is arising from only one ionisation state}. 
We find that $\dot{M}$ cannot be below $\sim\,$10$^{-8}$\,M$_{\odot}$\,yr$^{-1}$ and is
%if all oxygen in the line-forming region is in O$^{3+}$, or when summing the total of contributions of Ne$^{4+}$ and Ne$^{5+}$, 
instead in the range $\dot{M} = 3-5 \times 10^{-8}$ M$_{\odot}$ yr$^{-1}$ for $T_{\rm e}$$\sim$35 kK and
in the range $\dot{M} = 4-7 \times 10^{-8}$ M$_{\odot}$ yr$^{-1}$ for $T_{\rm e}$$\sim$200 kK.
These estimates are broadly within the range of theoretical mass-loss rates predicted by 
\cite{krticka17} and \cite{vink00}, and slightly larger than those predicted by \cite{bjorklund21}.\footnote{
A range of log$(L/L_{\odot})$ values have been quoted in the literature, with \cite{martins04} estimating a luminosity of 0.25 dex higher than our adopted value \citep{aschenbrenner23}; such a higher luminosity would correspond to a 0.55 dex increase in log$(\dot{M})$ for all predicted mass-loss rates.}

Since our observational estimates assume that the entire neon and oxygen mass budget is in the observed ions, any unobserved species would drive $q$ downwards, resulting in a higher total mass outflow rate.
%above the predictions of \cite{bjorklund21} for $T_{\rm e}$$\sim$35kK, close to the predictions from \cite{krticka17} and slightly below the predictions from \cite{vink00} for $T_{\rm e}$$\sim$35kK (and higher for $T_{\rm e}$$\sim$200kK). 
 Regardless of the exact temperature and luminosity, the mass-loss rate estimated from the IR spectra is significantly higher than those determined from previous optical and UV analyses, and suggest that forbidden lines in the IR are better probes of the hot, low density wind regime than traditional optical diagnostics.

\begin{figure}[t!]
    \includegraphics[width=\columnwidth]{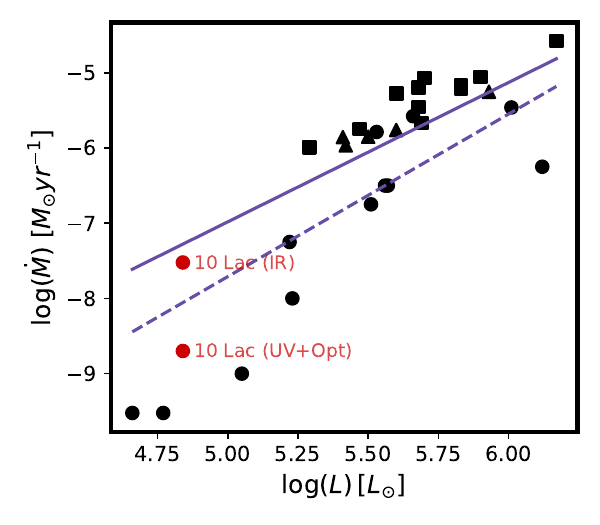}
    \caption{Mass-loss rates as a function of luminosity for Galactic O stars using results from \cite{repolust04, martins05, crowther06}. Circles, triangles and squares correspond to luminosity classes V, III and I respectively. Solid and dashed lines show theoretical mass-loss rate predictions from \cite{vink00} and \cite{bjorklund21} respectively.
    }
    \label{fig.mdot-logl}
\end{figure}

In Figure \ref{fig.mdot-logl} we compare measured mass-loss rates as a function of luminosity for a sample of Galactic O stars compiled from the literature, along with the theoretical predictions of \cite{vink00} and \cite{bjorklund21}. This comparison serves to highlight the discrepancy  -- the ``weak-wind problem" -- for stars 
with $\log(L/L_{\odot} ) \lesssim 5.2$.
Figure \ref{fig.mdot-logl} also shows that -- in stark contrast to previous determinations from optical and UV diagnostics -- the mass-loss rate we estimate for 10 Lac from the \nev, \nevi, and \ofour\ emission is in better agreement with theoretical predictions.
It would appear that the resolution of the ``weak-wind problem" for 10 Lac requires inclusion of a hot, low-density component that is difficult to detect with the diagnostics available in the optical and UV.

Hot, rarefied gas is a well-known by-product of the shocks produced by the potent line-deshadowing instability in line-driven stellar winds \citep{owocki88}.
This shock-heated gas is generally thought to be the origin of X-rays from non-magnetic massive stars \citep[see, e.g.,][]{feldmeier95}, and could also be responsible for the production of Ne$^{4+}$ and Ne$^{5+}$.
Although it is traditionally assumed that the dense, ``clumped" wind carries most of the mass, the current analysis for 10 Lac suggests that significant material also resides in the low-density, inter-clump medium.
In this connection, the two-component phenomenological model developed by \citet{lucy12} that assigns a larger volume to the hot, shocked gas than the cool, dense gas in the winds of low-luminosity stars merits consideration.

More observations of the mid-infrared fine-structure lines of highly ionized species with JWST/MIRI are clearly desirable to characterize the occurrence and behavior of these newly accessible diagnostics.
A particular aim is to determine whether the presence of hot, rarefied gas is a general solution to the ``weak-wind problem", or if 10 Lac is a special case.

Initial indications are that the situation may be complicated.  For example,
\citet{law24} also observed the ``weak-wind", O9.5 V star $\mu$~Columbae as a MIRI calibrator.  The MIRI spectrum of $\mu$~Col shows no evidence for the broad, high-ionization emission features seen in 10 Lac, despite the overall similarity of the fundamental stellar parameters of the two stars. Curiously, $\mu$~Col has a larger X-ray luminosity than 10~Lac \citep{berghoefer96}, which might suggest that it also has a wind with gas at even higher temperatures than in 10 Lac.
By analyzing X-ray profiles, \cite{huenemoerder12} determined a mass-loss rate for  $\mu$~Col that is consistent with theoretical predictions and much higher than what could be estimated from UV or optical diagnostics, thereby providing a potential solution to its weak-wind problem.   
%However, the same analysis indicates that the hot gas is near the star, at higher densities that might preclude the formation of fine-structure lines.  
The details of whether such winds are visible in the UV/optical, mid-infrared, or X-ray may therefore be intrinsically tied to the temperature and density profile of the winds and the most efficient radiation mechanisms available in each regime.
Nonetheless, it appears likely that further observations of the fine-structure lines in the mid-infrared for a carefully selected sample of early-type stars will help to provide qualitatively new diagnostics that can allow us to better disentangle the nature of winds around massive stars.

\vskip 20pt

%\begin{acknowledgments}
\noindent This work is based on observations made with the NASA/ESA/CSA James Webb Space Telescope. The data were obtained from the Mikulski Archive for Space Telescopes at the Space Telescope Science Institute, which is operated by the Association of Universities for Research in Astronomy, Inc., under NASA contract NAS 5-03127 for JWST. These observations are associated with programs \#1524 and \#3779 and can be accessed via \dataset[doi: 10.17909/s1xy-k738]{https://doi.org/10.17909/s1xy-k738}. 

%\end{acknowledgments}

\begin{appendix}

\section{Mass-loss rate calculation}
\label{appendix.sec}

We estimate the mass loss rate from the observed emission line flux by combining continuity
equations with our knowledge of the stellar atmospheric composition of 10 Lac.  The fractional ionic abundance $\gamma_i$ of a given element ($Z$) is defined as

\begin{equation}
\gamma_{i} \equiv \left(\frac{N_Z}{N_{\rm H} + N_{\rm He} + N_{\rm C} + N_{\rm N} + N_{\rm O} + ...}\right)
%\frac{N_{i}}{N_{H}} = \gamma_{i} \left(1 + \frac{N_{He}}{N_{H}} + \frac{N_{C}}{N_{H}} + \frac{N_{N}}{N_{H}} + \frac{N_{O}}{N_{H}} + ... \right)
\label{gamma1.eqn}
\end{equation}

%where $N_i$ is the number fraction in a given element, corresponding to a total mass fraction

%\begin{equation}
%X_{i} = \frac{N_{i}}{N_{\rm H}}\frac{m_{i}}{m_{\rm H}} X_{\rm H}
%\label{massfrac.eqn}
%\end{equation}

where $N_Z$ is the number fraction in a given element, corresponding to an ionisation fraction $q_{i}$ for a given element

\begin{equation}
q_{i} = \frac{\gamma_{i}}{\frac{N_{Z}}{N_{\rm H}} \mu X_{\rm H}} 
\label{massfrac.eqn}
\end{equation}

%where $m_{i}$ is the atomic mass of the element (e.g. $m_{\rm Ne}=20$, $m_{\rm O}=16$, and $m_{\rm H}=1$). We assume that oxygen ions in the O$^{3+}$ phase traced by \ofour\ emission are representative of the total mass fraction of oxygen ($\gamma_{\rm O} = \gamma_{\rm O^{3+}}$), and that neon ions in the Ne$^{4+}$ and Ne$^{5+}$ phases traced by \nev\ and \nevi\ respectively contribute equally to the total mass fraction of neon ($\gamma_{\rm Ne} = \gamma_{\rm Ne^{4+}} + \gamma_{\rm Ne^{5+}}$). 
where the total ionisation $q$ is the sum of all contributing ionisation fractions $q = \sum_{i} q_{i} = 1 $. $\mu$ is the mean ionic mass ($\mu = \sum\gamma_{i}m_{i}/\sum\gamma_{i}$) and is commonly around 1.5-1.6 for O-type stars. We adopt the photospheric abundances of each element for 10 Lac (e.g., $X_{\rm H} = 0.725$, $X_{\rm Ne} = 2.5\times10^{-3}$, $X_{\rm O} = 6.8\times10^{-3}$) from  \cite{aschenbrenner23}, and assume that these relative abundances remain constant throughout the surrounding stellar wind.

The fractional ionic abundance $\gamma_{i}$ can also be computed with an integral of the observed emission line flux over the line forming region surrounding the star (parameterized in terms of either the radius or the electron density) for a given atomic transition.  Assuming the case of a stellar wind following an $r^{-2}$ density law outflowing at fixed velocity $v_{\infty} = 1170$ \kms, equations 5 and 6 of \citet{crowther24}\footnote{See also full derivation by \citet{dessart00}.} together give

\begin{equation}
\gamma_{i} = \frac{(4\pi\mu m_{H} v_{\infty})^{1.5}}{ln(10)\dot{M}^{1.5}}  \frac{2d^{2}I_{ul}}{\sqrt{\gamma_{e}} A_{ul} h \nu_{ul}} \left( \int_{0}^{\infty} \frac{f_{u}(N_{e},T_{e})} {\sqrt{N_{e}}} \rm{d}\,log(N_{e})  \right)^{-1} 
\label{gamma2.eqn}
\end{equation}

where $\gamma_{e}$ is the mean number of electrons per ion which is commonly 0.8-1.0 for O-type stars (e.g. \citealp{leitherer95}). Likewise, $m_{H}$ is the mass of hydrogen, $h$ is Plancks constant, $A_{ul}$ is the Einstein coefficient (taken from NIST), and $\nu_{ul}$ is the frequency of the line transition. $f_{u}(N_{e},T_{e})$ is the fraction of ions in the upper level for the atomic transition in question and is calculated using the EQUIB code \citep{howarth81, danehkar20} in the range of $N_{e} = 10^{1} - 10^{12}$ for a given $T_{e}$, following \citet{dessart00} and \citet{crowther24}. $d$ is the distance to 10 Lac, which we take as 542~pc following \citet{aschenbrenner23}. 
%The clump volume filling factor modifies the wind density through $\rho = \bar{\rho}/f_{cl}$ where $\bar{\rho}$ is the mean wind density, therefore $f_{cl}$ can be varied in the range of 0 to 1 and is radius dependent. 

Combining equations \ref{gamma1.eqn} and \ref{gamma2.eqn}, it is therefore possible to infer a mass-loss rate $\dot{M}$ as a function of the line flux $I_{ul}$ in each observed \nev, \nevi, and \ofour\ transition.

We note that the most common simple clumping models for stellar winds 
\citep[as discussed by, e.g.,][]{dessart00,crowther24} assume that line emission arises in the cooler, overdense clumps.  This assumption is incompatible with emission from extremely highly ionized species such as \nevi\, which is more likely to be produced in the hot interclump medium instead \citep[e.g.,][]{lucy12}.  Such clumping models are therefore not directly applicable to our observations, and we instead assume a smooth wind profile, deferring detailed consideration of multi-component winds to future work.

\end{appendix}

\end{document}